\documentstyle[12pt,aaspp4,flushrt]{article}

\newcommand{\etal}{{\rm et al.~}}
\newcommand{\Mpc}{$h^{-1}$~{\rm Mpc}}
\newcommand{\hmpc}{$h$~{\rm Mpc$^{-1}$}}

\begin{document}

\title{Steps toward the power spectrum of matter.  
III. The primordial spectrum}

\author{Jaan Einasto \altaffilmark{1},
{Maret Einasto} \altaffilmark{1},
{Erik Tago} \altaffilmark{1},
{Alexei A. Starobinsky}\altaffilmark{2},
{Fernando Atrio-Barandela }\altaffilmark{3},
{Volker M\"uller}\altaffilmark{4},
{Alexander Knebe}\altaffilmark{4}, and
{Renyue Cen}\altaffilmark{5}}

\altaffiltext{1}{Tartu Observatory, EE-2444 T\~oravere, Estonia}
\altaffiltext{2}{Landau Institute for Theoretical Physics, Moscow 117334,
Russia}
\altaffiltext{3}{F{\'\i}sica Te\'orica, Universidad de Salamanca, 37008
Spain }
\altaffiltext{4}{Astrophysical Institute Potsdam, An der Sternwarte 16,
        D-14482 Potsdam, Germany}   
\altaffiltext{5}{Department of Astrophysical Sciences, Princeton University,
Princeton, NJ 08544, USA }

\begin{abstract}

We compare the observed power spectrum of matter found in Papers I and
II with analytical power spectra. We extrapolate the observed power
spectra on small scales to find the linear power spectrum of matter.
We consider spatially flat cold and mixed dark matter models with
cosmological constant as well as open models. We fix the Hubble
constant and the baryon density in the middle of the allowed range and
vary the density parameter and the cosmological constant. We determine
the primordial power spectrum of matter using the power spectrum of
matter and the transfer functions of analytical models.  We take two
different spectra suggested by observations: one with a sharp maximum
at 120$h^{-1}$Mpc and a second one with a broader maximum, as found
for regions with rich and medium rich superclusters of galaxies,
respectively. For both models, the primordial power spectra have a
break in amplitude; in the case of the spectrum with a sharp maximum
the break is sharp.  We conclude that a scale-free primordial power
spectrum is excluded {\em if} presently available data on the
distribution of clusters and galaxies represent the true mass
distribution of the Universe.

\end{abstract}



\section{Introduction}

Here we accept the current inflationary paradigm and assume that the
structure evolves from Gaussian initial conditions in a Universe
dominated by non-baryonic dark matter (DM). Under these assumptions
the current power spectrum of matter is determined by physical
processes during the inflationary expansion, and the subsequent
radiation- and matter-dominated eras.  It is believed that the
primordial power spectrum, formed during the inflation epoch, is a
scale-free Harrison-Zeldovich power law.  This scale-free power
spectrum is transformed during the radiation-dominated expansion, the
transformation depends on cosmological parameters and on the nature of
the DM. Within the presently acceptable range of cosmological parameters
and possible DM candidates, the transfer function (which describes the
transformation of the power spectrum) is a smooth function of scale.
For this reason the present power spectrum should be a smooth function
of scale.

However, recent evidence (Broadhurst \etal (1990), Landy \etal (1996),
Einasto \etal (1997a), Retzlaff \etal (1998), and Tadros \etal (1998))
indicates that the power spectra of galaxies and clusters of galaxies
have a spike or peak on scales around $l=120$~\Mpc\ or wavenumber
$k=2\pi/l=0.05$~\hmpc\ (we designate the Hubble constant as $100~h$
km~s$^{-1}$~Mpc$^{-1}$).  This scale corresponds to the step size of
the supercluster-void network (Einasto \etal 1994b, 1997c, hereafter
E97c).  Atrio-Barandela \etal (1997), and Eisenstein \etal (1998) have
shown that the peaked power spectrum of galaxies and clusters, and the
respective angular power spectrum of temperature anisotropy of the
cosmic microwave background (CMB) radiation are barely compatible with
the standard scale-free primordial power spectrum in a cold dark
matter (CDM) dominated Universe. An agreement is possible only for
extreme values for cosmological parameters (Hubble constant and baryon
fraction of the matter) which are almost outside the limits of the
range allowed by other data.

Motivated by the difficulty to reconcile the observed power spectra
with CDM-type models we try to calculate the primordial power spectrum
of matter empirically.  We shall use the latest determinations of the
power spectra for various populations of galaxies and clusters of
galaxies reduced in amplitude to the power spectrum of matter in the
local Universe (Einasto \etal 1999a, 1999b, hereafter Papers I and II,
respectively).  For comparison we use theoretical models with cold
dark matter (CDM), a mixture of cold and hot dark matter (MDM) in
spatially flat universe, as well as open models (OCDM). We calculate the
transfer functions for a set of cosmological parameters chosen in the
range of astrophysical interest, and find the primordial spectra from
theoretical transfer functions and empirical power spectra.  We
calculate for our empirical power spectra the mass function of
clusters of galaxies using the Press \& Schechter (1974) algorithm,
and compare them with the empirical cluster mass function.

\section{Comparison with theoretical models}

\subsection{Observed power spectra}

We shall use the mean galaxy power spectra determined in Paper I,
where we derived two mean power spectra based on different observed
populations. Clusters of galaxies and several galaxy surveys yield a 
mean power spectrum with a peak on a scale of 120~\Mpc. This spectrum,
$P_{HD}(k)$, characterizes the distribution of clusters and galaxies
in a large volume that includes rich superclusters of galaxies.  For
comparison we use also the power spectrum found for medium-density
regions of the Universe, $P_{MD}(k)$ (subscripts HD and MD denote
high- and medium-density regions).  This spectrum is shallower around
the maximum.  The power spectra are determined in the range $0.03 \le
k \le 1$~\hmpc.  On large and intermediate scales ($k \le 0.2$~\hmpc)
they are found on the basis of 3-dimensional galaxy and cluster
surveys.  Both mean power spectra coincide on small scales, where they
are based on the reconstruction of the 3-D spectrum from the 2-D
distribution of APM galaxies.  Spectra were reduced to real space and
to the amplitude of the power spectrum of all galaxies, including
dwarf galaxies.  This procedure assumes that there exists such a mean
power spectrum characteristic for all galaxies.  Possible errors of
this assumption and intermediate steps of the data reduction shall be
discussed below in Section 3.1.

Both mean power spectra were reduced in amplitude to the power spectra
of matter using the bias factor calculated from the amount of matter
in voids; the present epoch was fixed using the rms density
fluctuations in the 8~\Mpc\ sphere, $\sigma_8$ (Paper II).  The
biasing correction is based on the assumption that the structure
evolution in the universe on scales of interest is due to
gravitational forces alone.  Possible errors related to biasing
correction shall also be discussed below.

Observed power spectra are shown in Figure~1. They differ only on
scales around the maximum which is the most interesting part of the
spectrum for the present analysis.  On small scales, $k \ge
0.2$~\hmpc, the power spectra are influenced by non-linear effects.

\begin{figure}[ht]
\vspace*{7cm}  
\caption{Empirical power spectra of matter. $P_{HD}$ is the non-linear
power spectrum of matter with its $1\sigma$ error corridor (errors
were calculated from errors of observed power spectra);
$P_{HD-lin-MDM}$ and $P_{MD-lin-MDM}$ are the linear power spectra
extrapolated on small scales using the tilted MDM model with
$\Omega_0=0.4$, for observed spectra $P_{HD}(k)$ and $P_{MD}(k)$,
respectively; $P_{lin-CDM}$ is the linear power spectrum extrapolated
on small scales using the CDM model with $\Omega_0=0.4$. On large
scales both linear power spectra were extrapolated using the standard
CDM model with $\Omega_0=1$.  }
\includegraphics{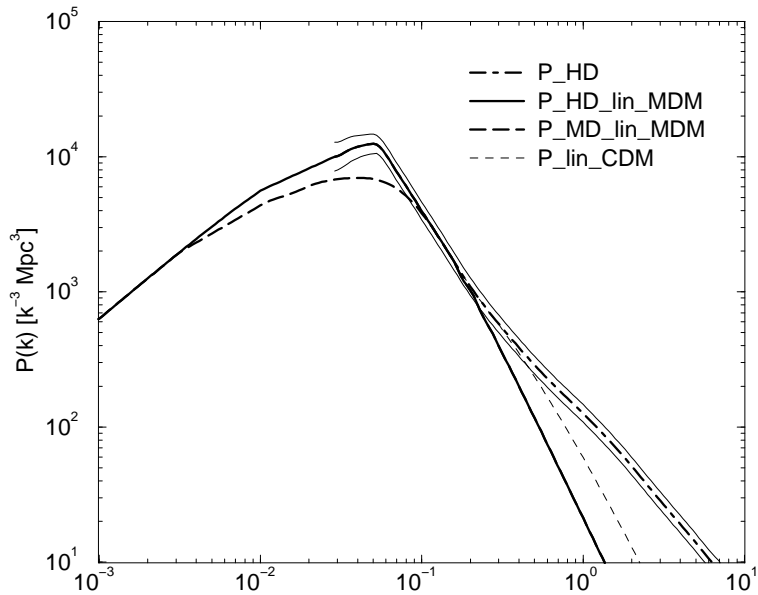}
\label{figure1}
\end{figure} 

\subsection{Extrapolation of power spectra on large and small scales}

To compare observed power spectra with theory they are to be reduced
to the linear case on small scales. To find the linear power spectrum,
Peacock (1997) used a method based on numerical simulations.  We shall
use a different method -- the comparison with theoretical power
spectra. Theoretical spectra shall be discussed in more detail in the
following Sections.  Spectra were calculated for CDM and MDM models
and for different values of the density parameter, $\Omega_0$; spectra
were COBE normalized.  The comparison shows that theoretical spectra,
calculated for high values of the density parameter, have amplitudes
much higher on small scales than the amplitude of observed power
spectra. The shape of theoretical power spectra based on the CDM model
is also rather different from the shape of observed power spectra on
scales between the maximum of the spectrum and the beginning of the
non-linear part of the spectrum at $k \approx 0.2$~\hmpc.  The best
fit, both of the amplitude (near the beginning of the non-linear
scale) and of the shape, is obtained with a tilted MDM model with a
power index of $n=1.1$ on large scales, and a density parameter
$\Omega_0=0.4$ (see Figure~1). For CDM models the best fit is obtained
(also of the amplitude and the shape) for a model with density
parameter $\Omega_0 \approx 0.2$; but now the fit is good in a much
narrower scale interval, see Figure~2 below.  For comparison we show
in Figure~1 an alternative linear extrapolation with a CDM model and a
density parameter, $\Omega_0=0.4$. In the following discussion we
shall use only the extrapolation with the MDM model.  Note that for
our choice of $h$ and $\Omega_{\nu}$ (for the MDM model), these
parameters corresponds to a neutrino rest mass (or the sum of the
masses of 2 or 3 stable neutrino species) of $\approx 3.4$ eV.

Now we consider the extrapolation to large scales.  The comparison of
observed power spectra with theoretical ones shows that all COBE
normalized models with low density parameter $\Omega_0$ have 
amplitudes which are higher on large scales than those of the observed
power spectrum (see Figure~2 below).  For this reason we have used the
standard CDM model with $\Omega_0=1$ to extrapolate the observed
spectra to large scales. This model has the lowest amplitude on large
scales and yields the best and smoothest extrapolation of observed
spectra on large scales. On these scales there are no essential
differences between theoretical spectra of CDM and MDM models.

We consider these extrapolations on small and large scales as
estimates of the true linear power spectrum of matter. In the
following we shall use the term ``empirical power spectrum'' to denote
this linear power spectrum. We recall that it is reduced to real
space, and its amplitude to the amplitude of the matter density
fluctuations.  As we have two observed mean power spectra of galaxies,
characteristic for different populations, we also have two empirical
power spectra.

\subsection{Theoretical models}

Previous studies (Atrio--Barandela \etal 1997, Eisenstein \etal 1998)
have shown that it is difficult to make models agree with optical and
CMB observations using conventional cosmological parameters and a
scale-free primordial power spectrum.  Here we shall use a different
approach.  We accept cosmological parameters in the middle of the
allowed region, and try to find possible restrictions to the
primordial (initial) power spectrum.

To calculate the analytical power spectrum we use a Hubble parameter
$h = 0.6$. This value is a compromise between the value $h=0.55 \pm
0.10$ suggested by Sandage \& Tammann (1997) and $h=0.75 \pm 0.06$
favored by Freedman (1997).  We adopt a density of matter in baryons
of $\Omega_{b} = 0.04$ (in units of the critical density). This
parameter is in agreement with recent nucleosynthesis results
$\Omega_{b} h^{2}= 0.01 - 0.02$ (Songaila, Wampler \& Cowie 1997,
Schramm \& Turner 1998, Olive 1997, Burles \& Tytler 1998, Steigman
1998, Turner 1998, see also Ostriker and Steinhardt 1995).

We consider models with various density parameter $\Omega_0=\Omega_b +
\Omega_{DM}$, as the density is presently the parameter which is
determined with less accuracy.  Local methods of the determination of
the mass-to-light ratio in clusters and groups of galaxies yield
$\Omega_0 \approx 0.2$ (Bahcall 1997). Methods sensitive to the global
value of the density parameter yield diverging values. Dekel, Burstein
\& White (1997), using the POTENT method, obtain a value close to
unity; another treatment of the peculiar velocity field yields a lower
value (Freudling \etal 1998).  The distant supernova project suggests
that the Universe is speeding up, i.e. that the vacuum energy
contributes about 60~\% of the critical density, and that the matter
density is about 40~\% of it, close to results obtained with local
methods (Perlmutter \etal 1998, Riess \etal 1998).  Indirect methods
based on the cluster abundance evolution favor a value of $\Omega_0 =
0.2 - 0.4$ (White, Efstathiou \& Frenk 1993, Bahcall, Fan \& Cen 1997,
Fan, Bahcall \& Cen 1997, Bahcall \& Fan 1998). Further evidence for a
low-density Universe comes from the analysis of the Lyman-alpha forest
by Weinberg \etal (1998), who found for a flat Universe
$\Omega_0=0.34^{+0.13}_{-0.09}$ (1$\sigma$ errors).  A recent analysis
of all available data by Turner (1998) favors $h=0.60~-~0.65$,
$\Omega_0 \approx 0.4$ and $\Omega_{\Lambda} \approx 0.6$.

To fix the amplitude of the analytical power spectrum on large scales
we use the four-year COBE normalization (Bunn \& White 1997). We cannot
use simultaneously the $\sigma_8$ normalization on small scales as
we do not know which model best fits  both normalizations. It is just
our goal to try to find a model which satisfies both normalizations.

We use three sets of models: spatially flat models with cold and mixed
dark matter, and open models with cold dark matter. In first two
models we use the cosmological constant. Its necessity is strongly
favored by recent data on cluster abundance and supernova explosions
at high redshifts (Bahcall \etal 1997, Perlmutter \etal 1998, Riess
\etal 1998).  We vary the density parameter between $\Omega_{0} = 0.2$
and $\Omega_{0} = 1$, and we choose the cold dark matter (CDM) content
to obtain a spatially flat model, $\Omega_b + \Omega_{DM} +
\Omega_{\Lambda}=1$. Models are calculated for a number of spectral
indices on large scales, $n=1.0, ~1.1, \dots ~1.4$.

We use a similar set of parameters for mixed dark matter (MDM) models
with the only difference that a hot dark matter component with density
parameter $\Omega_{\nu}=0.1$ was added. The density parameter of the
cold dark matter was decreased to get a spatially flat model.  We use
a small fraction of the hot dark matter for two reasons. First, most
direct and indirect methods suggest that $\Omega_0$ should be less
than unity. $\Omega_{\nu}$ should be chosen in agreement with the
relation $\Omega_{\nu}/\Omega_0 \sim 0.2$ required to get the correct
value of $\sigma_8$ (see, e.g., Pogosyan \& Starobinsky (1993), and
Pogosyan \& Starobinsky (1995) for the case of a tilted initial
spectrum with $n>1$).  Second, in order to build up the fine
filamentary structure of faint galaxies observed in supervoids
(Lindner \etal 1995), the fraction of matter in the cold dark matter
must be significantly larger than that of the hot component (Frisch
\etal 1995).

In the open models (OCDM) we used values of $\Omega_0=1.0, ~0.9, \dots
~0.3$ for the matter density.  The baryon density was fixed as in all
other models, and the density of the CDM was chosen appropriately.

Theoretical power spectra for our models are plotted in the left panels of 
Figure~2 together with the empirical linear power spectra of matter.
Models were calculated with the CMBFAST package of Seljak \& Zaldarriaga
(1996). We shall discuss the linear power spectra and the comparison of
model spectra with observations in subsequent sections below.

\begin{figure}[ht]
\vspace*{17.2cm}
\caption{The empirical linear power spectra of matter compared with
theoretical and primordial power spectra. Left: present power spectra;
right: primordial spectra; in upper panels for CDM models, in middle
panels for MDM models, and in lower panels for OCDM models. Solid bold
lines show the empirical linear matter power spectrum found for
regions including rich superclusters, $P_{HD}(k)$; dashed bold lines
show the empirical linear power spectrum of matter $P_{MD}(k)$ for
medium dense regions in the Universe.  Model spectra with
$\Omega_{0}=1.0, ~0.9, \dots ~0.2$ are plotted with thin lines; for
clarity the models with $\Omega_0 = 1.0$ and $\Omega_0 = 0.5$ are
drawn with dashed lines. In the case of MDM the model of lowest
density parameter is that with $\Omega_0=0.25$.  Only spectra with
spectral index $n=1$ on large scales are plotted.  Primordial power
spectra are shown for the peaked matter power spectrum, $P_{HD}(k)$;
they are divided by the scale-free spectrum, $P(k) \sim k$.  }
\includegraphics{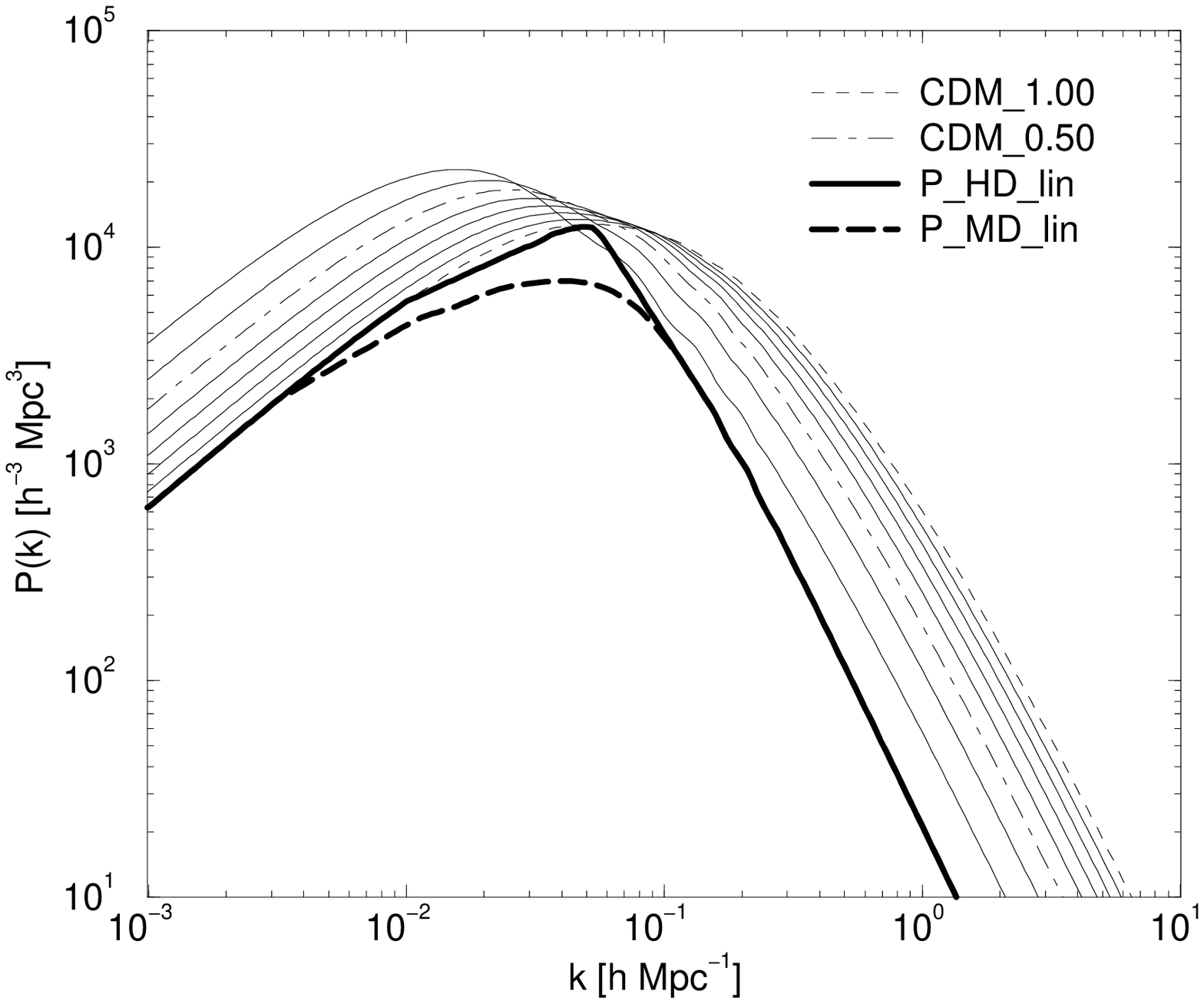}
\includegraphics{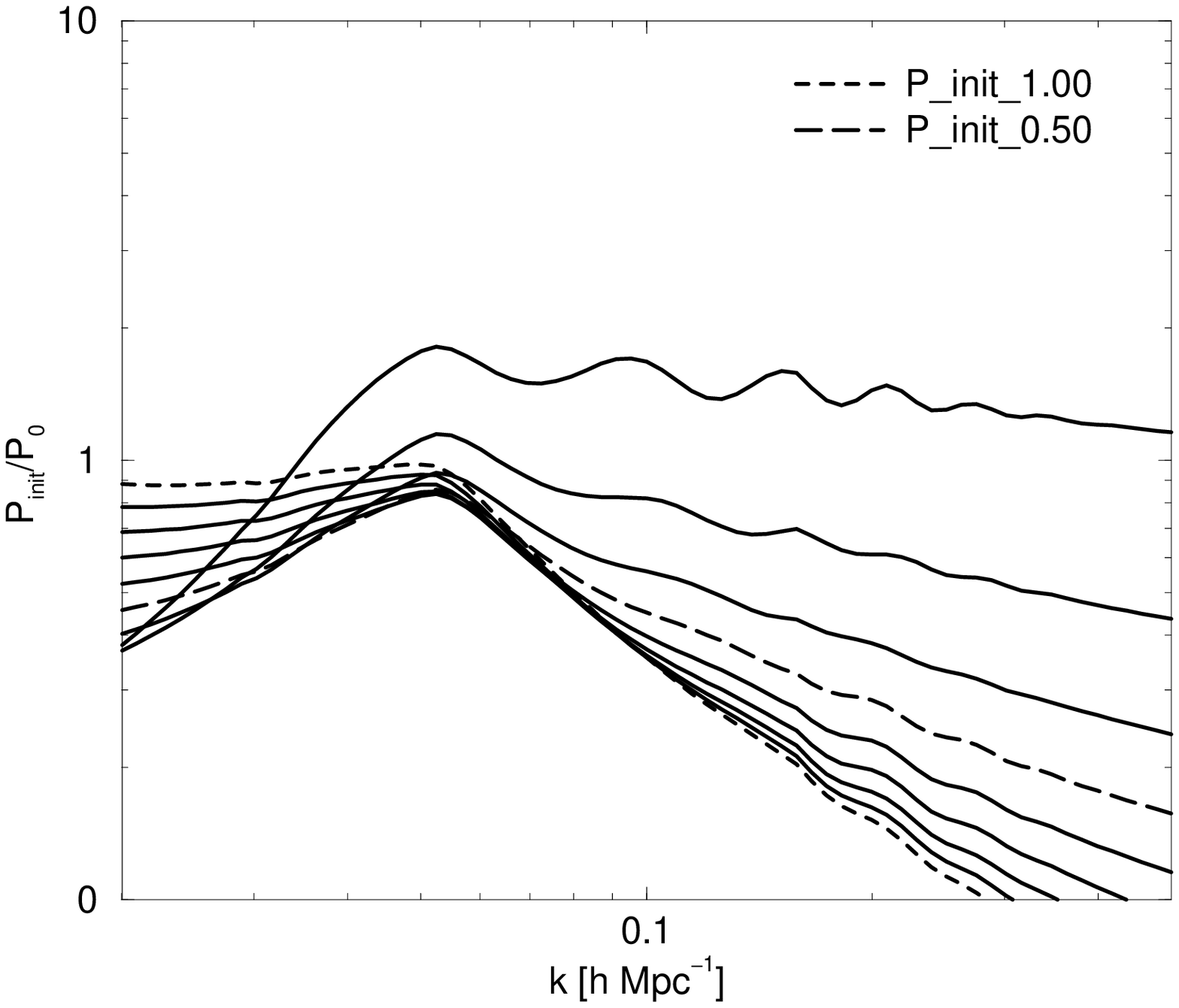}
\includegraphics{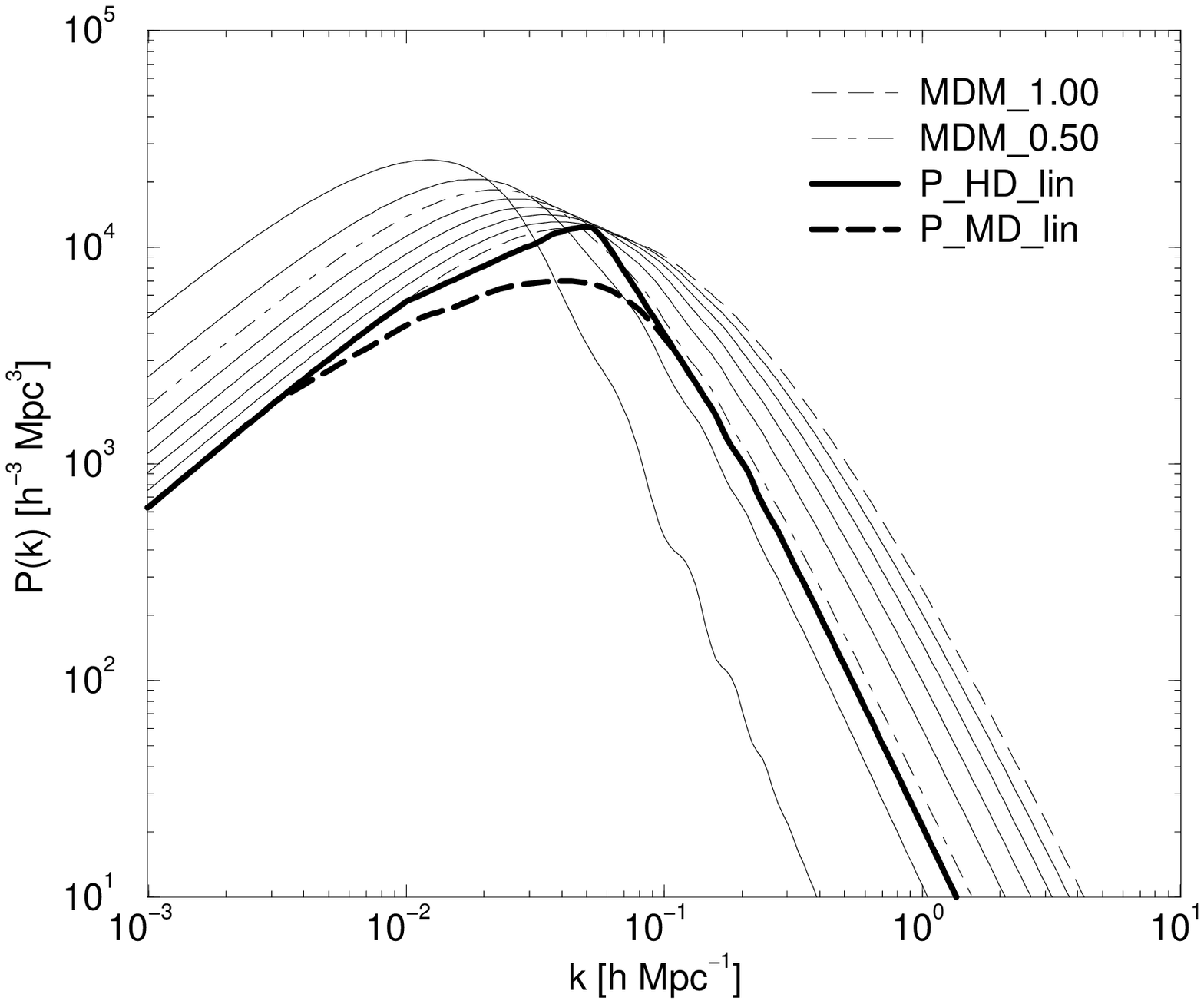}
\includegraphics{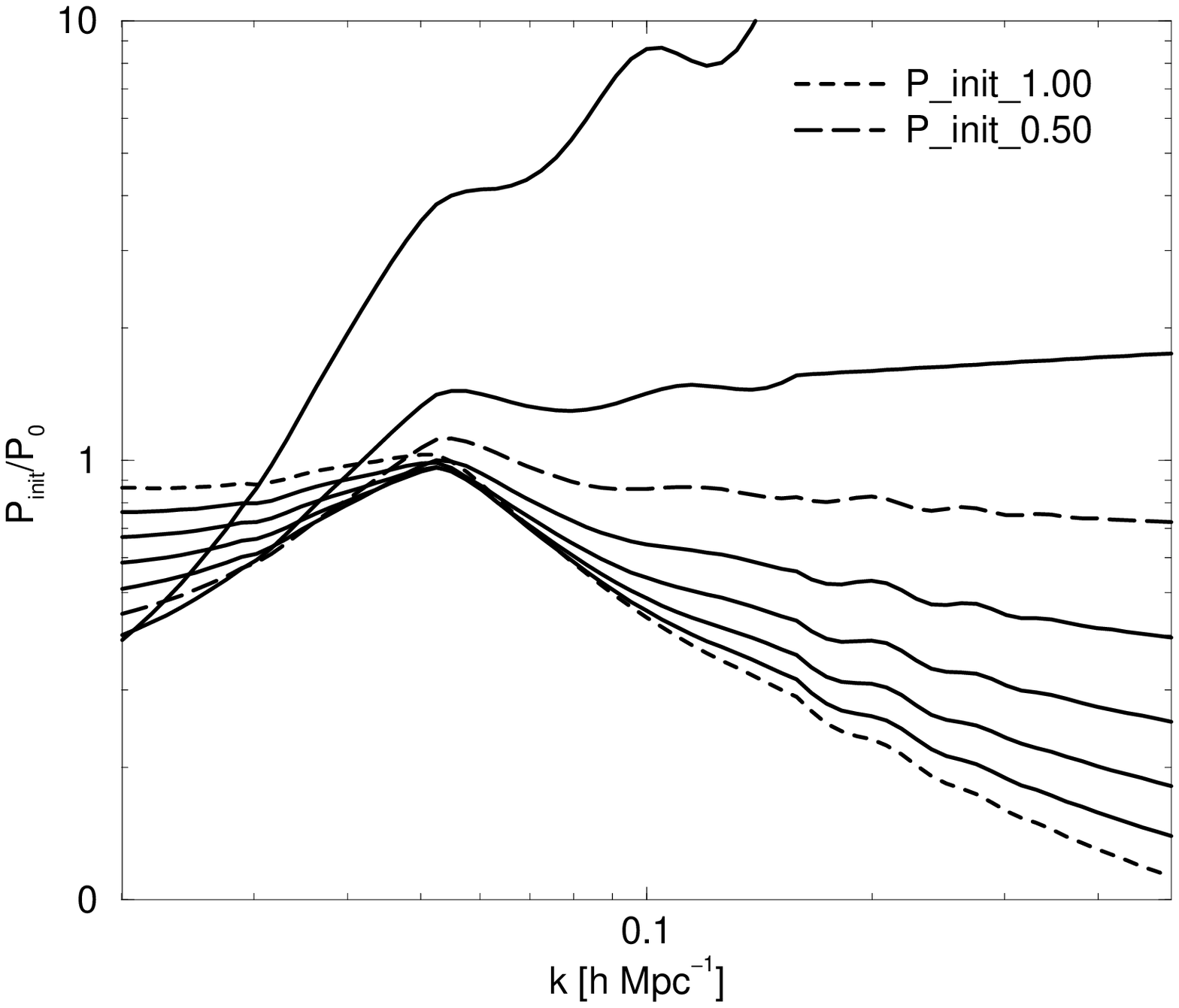}
\includegraphics{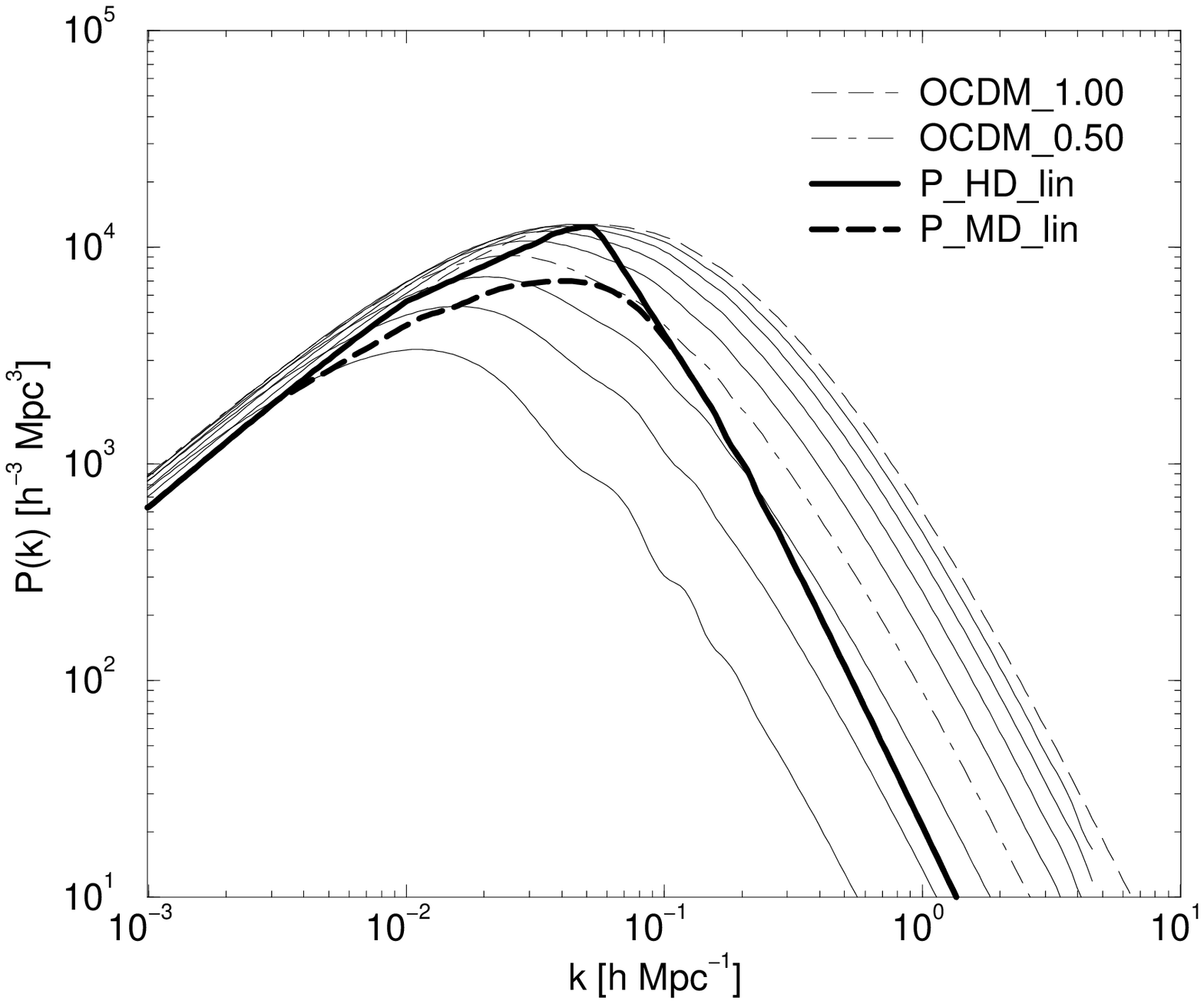}
\includegraphics{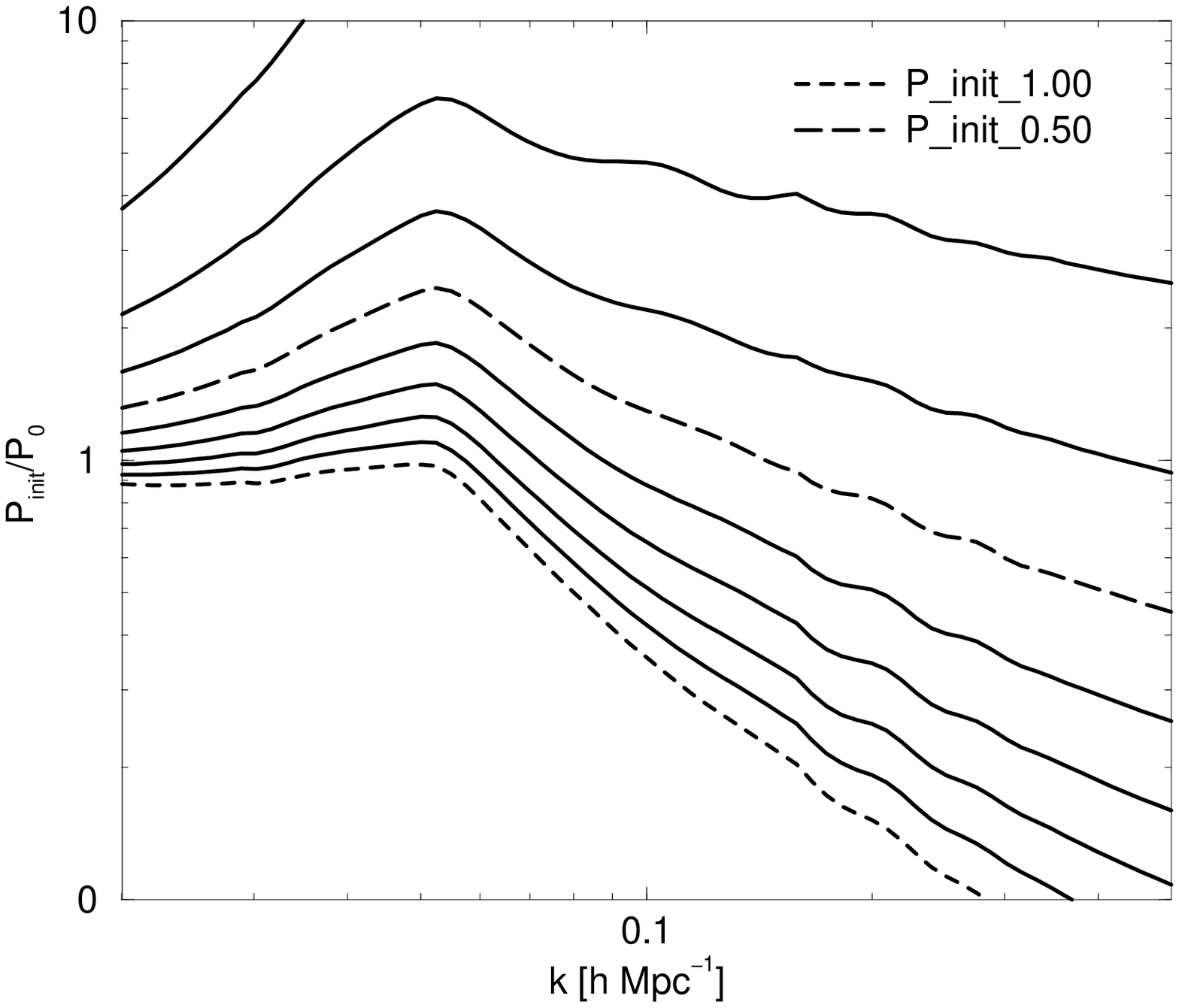}
\label{figure2}
\end{figure}

\subsection{Primordial power spectra}

The CMBFAST package yields the transfer function for each of our
models. We have used this function to calculate primordial power spectra,
$$ 
P_{init}(k) = P(k)/T^{2}(k),
\eqno(1)
$$ 
where $T(k)$ is the transfer function.  In the right-hand panels of
Figure~2 we plot the ratio of the primordial power spectrum to the
scale-free primordial power spectrum, $P_{init}(k)/ P_0(k)$, where
$P_0(k) \sim k$.  We plot here only results found for the peaked
empirical power spectrum $P_{HD}(k)$. A comparison for the flatter
power spectrum $P_{MD}(k)$ (which represents medium rich regions) is
given in Figure~3. In both cases we have used the empirical power
spectra extrapolated on small and large scales as described above.
The discussion of primordial power spectra follows in Section 3.3.

\section{Discussion}

\subsection{Power spectra}

We compare now empirical power spectra with theoretical ones on large,
intermediate and small scales.

On large scales the amplitude of COBE normalized theoretical power
spectra depends on the density parameter.  Low-density CDM and MDM
models dominated by the cosmological constant have amplitudes higher
than models of critical density.  Figure~2 demonstrates that on large
scales empirical power spectra have amplitudes which are compatible
with amplitudes of COBE normalized theoretical power spectra only for
high-density CDM and MDM models with $\Omega_0 \approx 1$.  For this
reason a smooth extrapolation of empirical power spectra on large
scales was possible only for high-density models.  The highest
amplitude on large scales (for extrapolation of observed power
spectra) which is still in satisfactory agreement with the COBE
normalization of theoretical spectra and with the upper limit of the
error corridor of empirical spectra, is the CDM or MDM spectrum with
$\Omega_{0}=0.7$ (the CDM and MDM model spectra are similar in this
range of scales).

\begin{figure}[ht]
\vspace*{6.5cm}
\caption{ Primordial power spectra for the empirical linear matter
power spectrum, $P_{MD}(k)$; divided by the scale-free spectrum, $P(k)
\sim k$. Left: CDM model with decreasing density parameter $\Omega_0$
from below on the right side of the panel; right: MDM with decreasing
$\Omega_0$ from below on the right side of the panel. Models with
$\Omega_0=1.0$ and $\Omega_0=0.5$ are plotted with dashed lines for
clarity.  }
\includegraphics{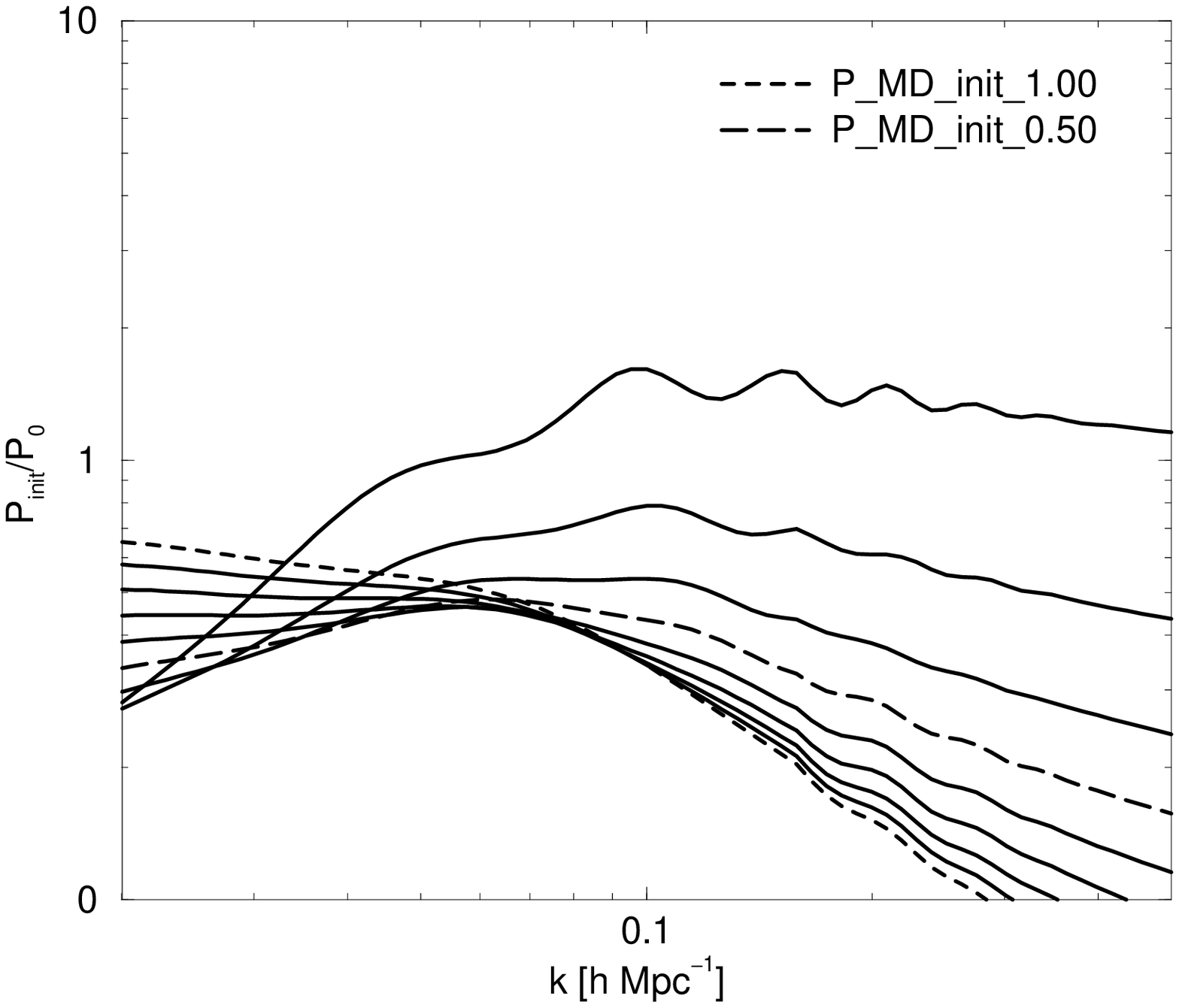}
\includegraphics{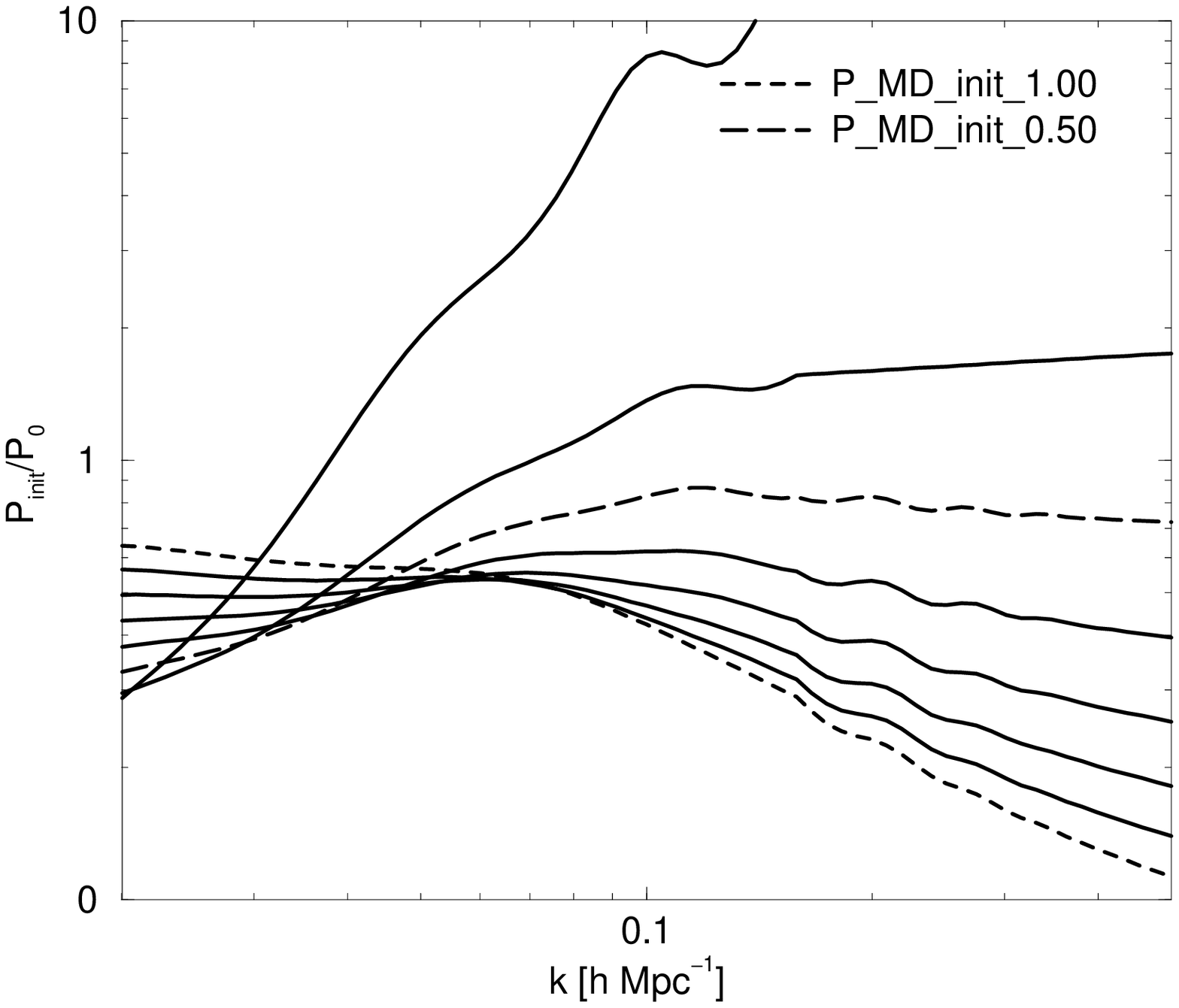}
\label{figure3}
\end{figure}

On intermediate and small scales, $k \geq 0.05$~\hmpc, the amplitude
of theoretical power spectra of CDM and MDM models varies considerably
with the density parameter. Only theoretical power spectra with a low
density parameter have amplitudes similar to the amplitude of the
empirical power spectra.  For CDM model the best agreement is achieved
for a model with $\Omega_0 \approx 0.2$, and for an MDM model for
$\Omega_0 \approx 0.4$ (see Figure~2). On these scales CDM and MDM
models are in agreement with other evidence, which suggests that
high-density models with $\Omega_0 \approx 1 $ are excluded by a large
margin (Ostriker and Steinhardt 1995, Bahcall, Fan, Cen 1997, Fan,
Bahcall, Cen 1997, Borgani \etal 1997, Bahcall \& Fan 1998, Turner
1998, Weinberg \etal 1998).  Low-density theoretical power spectra
have, however, maxima at $k \leq 0.01$~\hmpc\ with an amplitude of the
spectrum which is much higher than that of the observed peak at
$k=0.05$~\hmpc.  Available data do not support the presence of a
rising spectrum on these scales: spectra found from cluster and galaxy
data decrease in amplitude toward large scales.

Open models have a smaller increase of the amplitude on large scales
for a wide range of the density parameter, and are thus in a better
agreement with a reasonable extrapolation of the observed power
spectrum on large scales.  This agreement is lost around the maximum
of the empirical power spectrum, and both versions of empirical power
spectra deviate here from OCDM models by a large margin.

This comparison shows that it is impossible to satisfy the shape of
the empirical power spectrum with models with a fixed density
parameter simultaneously on large and small scales.  Models which fit
empirical spectra on large scales are incompatible with empirical
spectra on small scales and vice versa.  This is the main conclusion
obtained from the comparison of cosmological models with the data.

\subsection{Primordial power spectra}

Now we discuss primordial spectra derived from the empirical power
spectrum, $P_{HD}(k)$. On large scales there is little difference
between CDM and MDM models. The main feature of primordial power
spectra is the presence of a spike at the same scale as that of the
maximum of the empirical power spectrum.  On scales shorter than that
of the spike, the primordial spectrum can be well approximated by a
power law (a straight line in a log--log plot).  The slope of this
line varies with the cosmological density parameter accepted for
theoretical models. For models with high cosmological density,
$\Omega_0 \geq 0.5$, the slope is negative, $n<1$; for models with low
density, $\Omega_0 <0.5$, it is positive, $n>1$.  Such tilted
primordial power spectra (i.e. spectra with $n \neq 1$) have been used
to increase the accuracy of the model.  This approximation breaks
down, if we consider the whole scale interval.  On large scales the
primordial power spectrum can also be approximated by tilted
models. The power index of the approximation is, however, completely
different from the index suitable on small scales, as seen in
Figure~2.

Additionally, there is a considerable difference in the amplitude of
the primordial power spectrum on small and large scales.  For most
values of the cosmological density parameter the amplitude on small
scales is lower than on large scales (compared to the scale-free
primordial spectrum). For low values of $\Omega_0$ the effect has the
opposite sign: the amplitude of the primordial spectrum on small
scales is higher than on large scales. Such an effect has been 
noticed already by Lesgourgues \etal (1998).  The amplitude of the break
depends on the model used, and in MDM models it is larger than in CDM
models.  The primordial power spectrum for open model has similar
features, only its amplitude on large scales rapidly increases if we
consider more open models. This is due to the fact that the amplitude
of theoretical power spectra of these models on scales $k \geq
0.05$~\hmpc\ is much lower than the amplitude of the empirical power
spectrum.

Primordial power spectra, which are found from the comparison of the
flat empirical power spectrum, $P_{MD}(k)$, with theoretical spectra,
are plotted in Figure~3.  On very large and small scales both
empirical power spectra are identical, and consequently primordial
power spectra are identical, too.  Differences exist in the medium
scale range. Here primordial power spectra have a smoother transition
from long to small scales with no sharp spike at wavenumber $k=
0.05$~\hmpc. However, a change of the spectral index and amplitude
around this wavenumber (a ``break'') is very well seen.  This property
of the primordial power spectrum is similar for spectra derived on the
basis of CDM, MDM and OCDM models. An independent analysis of the
initial power spectrum was made by Adams, Ross \& Sarkar (1997) using
the APM galaxy power spectrum (which is identical to our $P_{MD}(k)$).
The presentation of the initial power spectrum (their Fig.~1) is
slightly different from ours, the main features of the spectrum are,
however, well seen, with similar conclusions.

Previous studies have shown that it is practically impossible to bring
the classical models into agreement with new data on the power
spectrum by varying cosmological parameters.  Our analysis confirms
these results. The reason for the discrepancy lies in the shape of the
empirical power spectrum: it is much narrower than theoretical
spectra.  A change of the bias parameter or a normalization on large
scales does not change these conclusions: they would only shift the
features present in Figures 2 and 3 up or down, without changing
considerably the shape of the primordial spectrum.

\begin{figure}[ht]
\vspace*{6.5cm}
\caption{ Power spectra (left) and correlation functions (right) of
MDM and OCDM models with density parameter $\Omega_0=0.4$, compared
with the linear empirical power spectrum of matter and correlation function of
clusters of galaxies in rich superclusters. The power spectrum of the
MDM model is calculated with spectral index $n=1.1$. Cluster
correlation functions are calculated via Fourier transform from power
spectra of matter, and are enhanced in amplitude by a biasing factor 7.7
which corresponds to the mean difference between respective power
spectra. The observed cluster correlation function is the one for
Abell-ACO clusters in very rich superclusters as derived by E97b.
}
\includegraphics{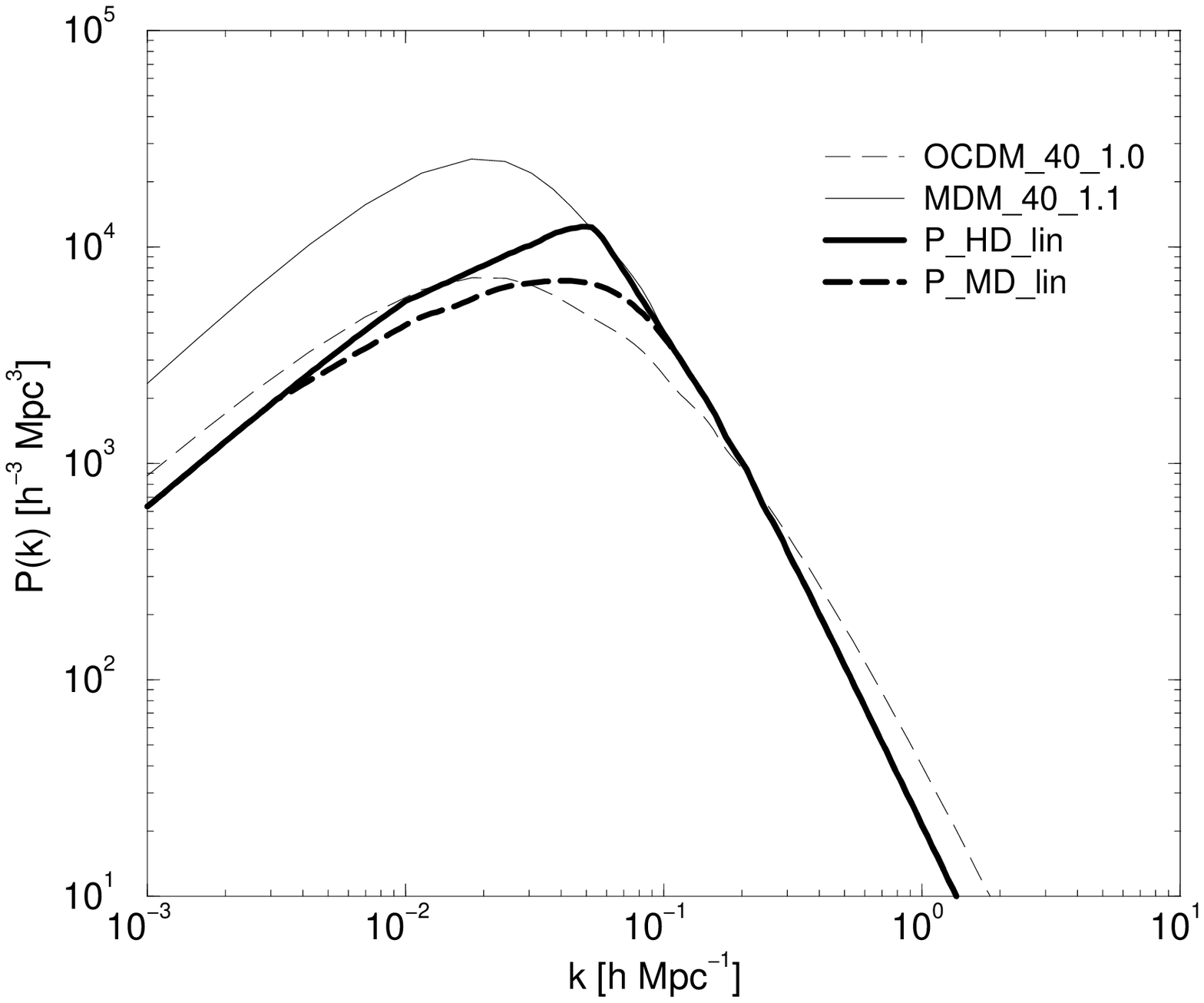}
\includegraphics{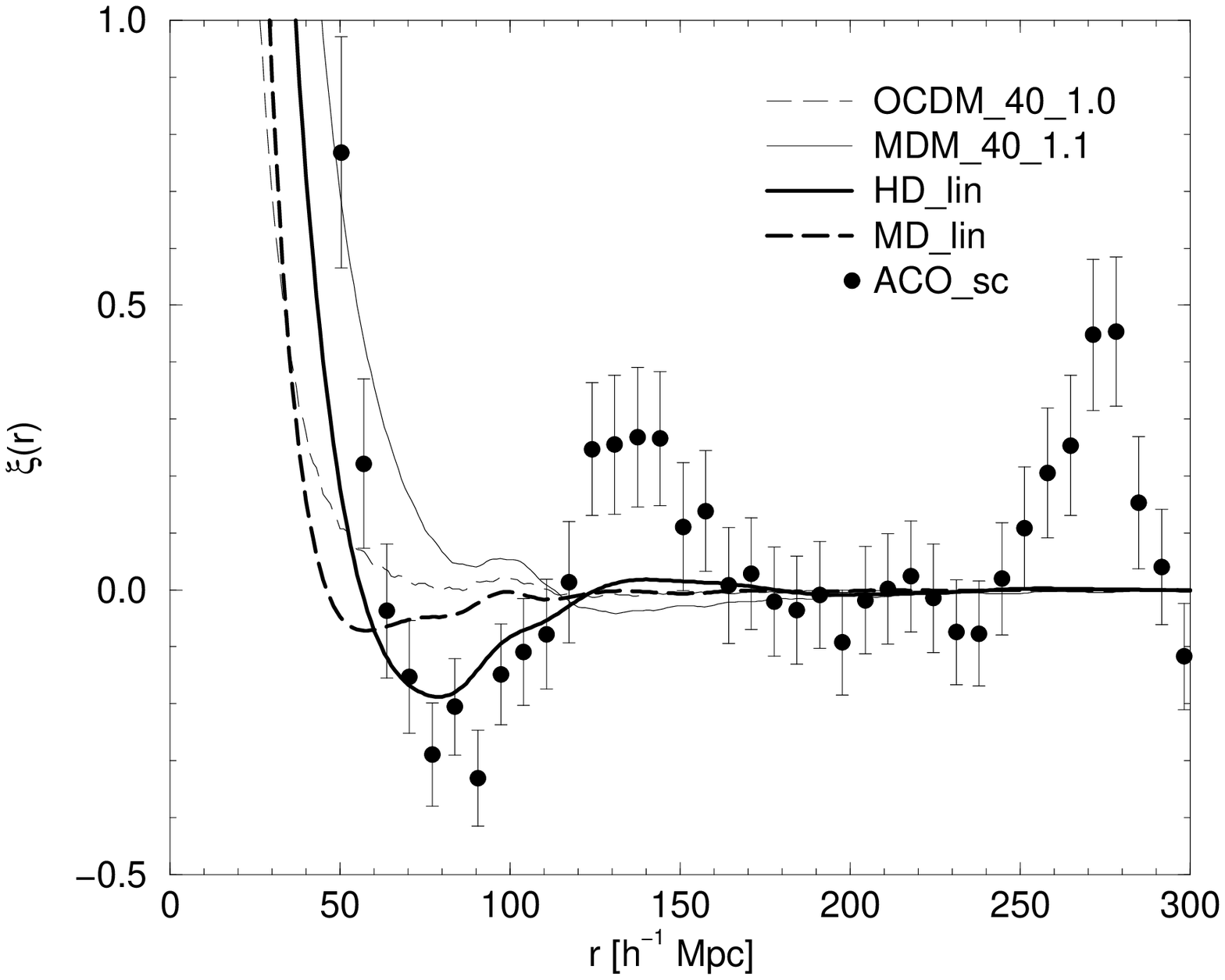}
\label{figure4}
\end{figure} 

\subsection{Models with low density parameter}

The previous analysis has shown that on scales $k \geq 0.05$~\hmpc\
the empirical power spectrum can be well approximated by a MDM model
with density parameter $\Omega_0 \approx 0.4$.  A slightly less
accurate approximation in a narrower scale interval is provided by a
CDM model with $\Omega_0 \approx 0.2$. An open model with density
parameter $\Omega_0 \approx 0.4$ gives an approximation which fits the
empirical power spectrum $P_{MD}$ both on large and small scales (but
not on intermediate scales). A spike or break of the primordial
spectrum at $k=0.05$~\hmpc\ could be avoided using one of these
models, if we allow for the higher amplitudes on large scales of the
empirical power spectrum (for CDM and MDM models), or a shallower
turnover on medium scales (for the OCDM model). Examples of such
spectra for MDM and OCDM models are shown in Figure~4.  It is
reasonable to investigate the possibilities of such models.

There are two major reasons to reject these model power spectra.
First, the directly observed power spectrum does not have a rising
section above the maximum at $k=k_0=0.05$~\hmpc, needed to fit
low-density CDM or MDM models, or a very shallow maximum needed for
the low-density OCDM model.  Second, as shown by numerical experiments
discussed in Paper I, models with rising or shallow power spectra in
this scale interval have a distribution of superclusters which is very
different from the distribution of real superclusters. In a model with
a broad power spectrum, such as all scale-free CDM, MDM and OCDM
models, the distribution of rich superclusters is close to a random
one.  Real rich superclusters form a quasi-regular lattice, which
surrounds large voids of diameter $\approx 120$~\Mpc\ (E97c). A
quantitative measure of the regularity is the cluster correlation
function on large scales, which oscillates with a period that
corresponds to the step size of the lattice (Einasto \etal 1997b,
E97b).

To test model power spectra for the regularity of superclusters we
have calculated via Fourier transformation correlation functions for
models shown in Figure~4. To compare with the correlation function of
clusters of galaxies in rich superclusters we multiplied correlation
functions with factors which correspond to the mean difference in the
amplitude of power spectra of matter and clusters in rich
superclusters. Correlation functions are plotted in the right panel of
Figure~4.  We see that the correlation function calculated from the
power spectrum $P_{HD}$ is oscillating with a low amplitude, both MDM
and OCDM models do not show signs of oscillations. We note that the
amplitude of oscillations of the observed cluster correlation function
is exaggerated due to the elongated form of the observed cluster
sample. However, as shown by simulations, it is impossible to generate
regular oscillations of the cluster correlation function from a random
distribution of superclusters.

To estimate the statistical significance of the difference between
observed and model power spectra we can apply the procedure used by
E97b.  The broad power spectrum of MDM and OCDM models can be
approximated by a geometrical model with randomly distributed
superclusters. E97b generated 1000 simulated catalogs of clusters,
applied selection criteria similar to those used in real cluster
samples, selected very rich superclusters, and calculated the
respective correlation functions. Very rich superclusters cause peaks
and valleys in the correlation function; the oscillatory behavior of
the correlation function was characterized by the period and amplitude
of oscillations, their rms scatter, and by a parameter which
characterizes the regularity of oscillations. This test shows that
some individual combinations of oscillation parameters are in
agreement with respective parameters of the real cluster correlation
function with a probability of about 1~\%, but a simultaneous
occurrence of all parameters has a much smaller probability (of the
order of $10^{-6}$). This test shows that there is a statistically
highly significant difference between samples which generate broad and
peaked power spectra.  A similar conclusion has been obtained by Luo
\& Vishniac (1993) by the comparison of the 1-D power spectrum found
for the Broadhurst \etal (1990) pencil beam survey with the
theoretical spectrum calculated on the basis of the standard theory.
Significant differences between cluster power spectra and spectra
based on various variants of CDM models have been noticed by Retzlaff
\etal (1998) and Tadros \etal (1998); similar differences have been
found also in case of the power spectrum based on the 2-D distribution
of APM galaxies by Peacock (1997) and Gazta\~naga \& Baugh (1998).

\subsection{Theoretical implications}

A consequence of the broken-scale-invariant (BSI) primordial power
spectrum is that one cannot draw conclusions based on the exact
scale-free primordial spectrum, such as the use of the zero crossing
of the correlation function, to determine the density parameter through
the parameter $\Gamma=\Omega_0 h$. Another conclusion is that one
should not expect the slopes $n_S$ and $n_T$ of power spectra of
primordial matter and gravitational waves to be independent on $k$,
especially, around the critical point $k=0.05$~\hmpc. Here $n_S$ and
$n_T$ correspond to matter (scalar perturbations) and gravitational
waves (tensor perturbations), respectively. In the latter region,
differential relations between these quantities following the
slow-roll approximation during inflation may become invalid.

Also, we have to understand the reason why the primordial power
spectrum deviates from the classical Harrison-Zeldovich spectrum. A
very general way to obtain such a behavior requires some kind of phase
transition which occured during inflation, about 60 e-folds before the
end of the inflation era. There is one specific model of this type
which has an exact solution. It was suggested by Starobinsky (1992),
and leads to a definite BSI spectrum.  It was recently confronted with
observational data by Lesgourgues \etal (1998).  However, there exist
a number of other possibilities (see recent discussions by Adams, Ross
\& Sarkar (1997) and Starobinsky (1998)) which should be explored,
too.

\section{Conclusions}

The main goal of this series of papers was to determine the mean
empirical power spectrum of galaxies, to reduce it to the matter, and
to compare with theoretical models to see whether it is possible to
fit empirical spectrum with models.  This procedure rests on several
assumptions and conclusions based on these assumptions. The main
assumptions are:

{$(1)$} there exists a power spectrum which represents all
galaxies in a fair sample of the Universe (including faint dwarf
galaxies); 

{$(2)$} the dynamical evolution of the Universe on scales of
interest is determined by gravity only; 

{$(3)$} density perturbations grow from small random fluctuation
generated in the early stage of the evolution;

$(4)$ the dynamics of the Universe is dominated by cold dark
matter with some possible mixture of hot dark matter; 
 
$(5)$ galaxy samples of various environment, morphology and
luminosity can be approximated by particles in numerical simulations
chosen in certain threshold density intervals.

The first assumption is based on observation of power spectra of
galaxy samples with different absolute magnitude limits: spectra are
identical in shape and amplitude if the sample contains sufficiently
faint galaxies (GE92).  The scatter of observed data points of power
spectra of faint galaxy samples is about 10~\% which can be attributed
to the cosmic scatter.  Thus we can expect that the assumption itself
does not introduce any systematic error to the analysis.  The last
assumption is based on our experience in comparing real and simulated
galaxies, collected in last ten years; its justification was analyzed
in detail in Paper II of this series.

The second, third and fourth assumptions form main paradigms of modern
cosmology.  The second and third assumptions are critical in the whole
chain of reduction steps of observed power spectra. Under these
assumptions, due to gravitational instability, the structure evolution
in high- and low-density regions is different: in low-density regions
matter remains in the primordial form, whereas in high-density regions
matter collapses and forms galaxies and systems of galaxies.  A clump
of matter collapses if its density is high enough, thus the
gravitational character of the evolution leads to the conclusion that
galaxy formation is a threshold phenomenon -- there exists a critical
threshold density, needed for a clump of matter to collapse.  Here the
{\em actual} density of the primordial matter is important.  Groups
and clusters of galaxies have a characteristic scale of the order of
1~\Mpc, thus calculated densities must not be smoothed over scales
exceeding this value. Various galaxy populations can be simulated
using different threshold densities (5th assumption). A threshold
density, equal to the mean density of matter, divides the population
of all galaxies from that of matter in voids, higher threshold
densities correspond to luminous galaxies, and to galaxies in
clusters.

The division of matter into unclustered and clustered populations
affects the respective power spectra in a simple way, so that power
spectra of galaxies, systems of galaxies, and of matter are similar in
shape.  We can reduce spectra of different populations to the power
spectrum of all galaxies by a shift in the amplitude only.  Simulation
show that relative errors of this reduction procedure are of the order
of 1~\% for galaxies and about 5~\% for clusters.

The comparison of power spectra of simulated galaxy and cluster
samples in real and redshift space has shown that the influence of
peculiar velocities in groups and clusters is serious on small scales
only. To avoid complications with the reduction of galaxy power
spectra to real space, we have accepted on small scales ($k \geq
0.1$~\hmpc) the power spectrum of APM galaxies as the power spectrum
of all galaxies in real space.  This power spectrum is deduced from
2-D distribution of galaxies, and has no redshift distortions.  Here
we assume that the APM sample is a fair sample of all galaxies in the
Universe on these scales. The main error of the amplitude of the power
spectrum of all galaxies is related with this assumption. The
amplitude may have a relative error of the order of 15~\%, which is
the largest possible error of the power spectrum of all galaxies.

On large scales the mean power spectrum is determined essentially by
clusters of galaxies.  Numerical simulation confirm that on these
scales the redshift correction (due to the contraction of
superclusters) affects only the amplitude of power spectra; possible
systematic errors involved are negligible in comparison with sampling
errors.

The final step in deriving the empirical power spectrum of matter is
the reduction of the mean power spectrum of all galaxies to that of
matter (the conventional biasing correction).  Here we use again our
basic assumptions that the structure evolution is due to gravity
alone, and that initial density fluctuation have a Gaussian
distribution. Additionally we use our last assumption that galaxy
populations can be approximated by samples of simulated particles
chosen in various threshold density intervals.  As discussed above,
under these assumptions the matter power spectrum can be found by
shifting the galaxy spectrum in amplitude.  The shift is determined by
the fraction of matter in the clustered population.  There are two
sources of error in this procedure: the uncertainty of the threshold
density level, $\varrho_0$, which divides the low-density population
of primordial particles in voids from the clustered population of
particles associated with galaxies; and errors due to uncertainty of
cosmological parameters which define the speed of void evacuation and
the fraction of matter in high-density regions (see Paper II for
details).  The relative error of the threshold density level was
estimated by Einasto \etal (1994a) to be 10~\%, which leads to a 3~\%
error of the fraction of matter in the clustered population and
respective biasing parameter. The second error can be estimated from
data given in Paper II (see Table~2 and Figure~4), it is 7~\%.
Additional sources of error are the fuzziness of the threshold
density, $\varrho_0$, and the assumption that the distribution of
luminous galaxies in superclusters, clusters and groups is similar to
the distribution of all matter.  The analysis done in Paper II shows
that these errors do not influence the shape of the power spectrum in
the scale range of interest (corresponding errors are of the order of
1~\%); the errors in the amplitude are also small, less than or equal
to 5~\%.  The total rms relative error of the biasing factor is 10~\%
(all errors are $1\sigma$ errors).

To summarize the overview of the data reduction procedure we can say
that the gravitational character of the structure evolution implies
that intermediate steps of data reduction introduce no noticeable
error to the shape of the power spectrum, possible errors are in the
amplitude of the power spectrum only.  The total rms error of the
amplitude is essentially determined by two errors: the error of using
the amplitude of the APM galaxy power spectrum as the amplitude of the
power spectrum of all galaxies in a fair sample of the Universe; and
the error due to the uncertainty of model parameters used in the reduction
of the galaxy power spectrum to matter.
  
Possible errors of the shape of the power spectrum are intrinsic, due
to differences of power spectra of different populations. To quantify
possible differences in the shape of the power spectrum of matter we
have formed two mean power spectra, characteristic for samples which
cover regions of the Universe including high-density superclusters,
and medium-density superclusters only.  Differences between power
spectra of these samples may partly be due to unknown errors of data
reduction used in various samples, or to the geometry of samples (thin
slices in case of the LCRS).

Empirical mean power spectra of matter are extrapolated on small
scales (to get linear power spectra) and on large scales (to have the
spectrum defined on all scales for comparison with theory). Power
spectra found for high- and medium-density regions are identical on
small scales ($k \geq 0.1$~\hmpc), and on large scales ($k <
0.01$~\hmpc).  On medium scales they are different. The power spectrum
found for regions which include high-density systems with rich
superclusters has a peak at $k = 0.05$~\hmpc, and an almost constant
power index $n= -1.9$ for scales in the range $0.05 < k < 0.2$~\hmpc.  The
power spectrum of medium-density regions has a shallower shape around
the maximum, the amplitude near the maximum is lower by a factor of 2.

Empirical power spectra can be well approximated by theoretical model
spectra based on various dark matter models in a limited scale range.
The best approximation for the empirical power spectrum for
high-density regions on medium and small scales is provided by a
spatially flat MDM model with cosmological density parameter
$\Omega_0=0.4$, baryonic density $\Omega_b=0.04$, cold dark matter
density $\Omega_{CDM} = 0.26$, hot dark matter density
$\Omega_{nu}=0.1$, and cosmological constant term $\Omega_{\Lambda} =
0.6$; Hubble parameter was fixed at $h=0.6$.  This approximation breaks
down on large scales: the model power spectrum continues to rise on
scales $k < 0.05$~\hmpc, whereas the amplitude of the empirical power
spectrum decreases.  The impossibility to represent empirical power
spectra with one theoretical model is one of the main results of this
series of papers.

If we make use of our fourth assumption that the dynamical evolution
of the Universe is dominated by some sort of dark matter, then we can
use our empirical power spectra to calculate the primordial power
spectrum.  These calculations show that for both variants of the
empirical power spectra the primordial power spectrum has a break,
i.e. it can be approximated on large and small scales by two power
laws with different power indices.  This is due to the fact that it is
impossible to approximate the empirical power spectrum with one single
model.  The strength of the break depends on the cosmological
parameters used for theoretical models, and the shape of the break is
different for empirical power spectra found for high- and
medium-density regions.

Our conclusions are based on the tacit assumption that galaxy and
cluster samples used represent the true matter distribution in the
Universe.  To check this assumption new deeper redshift data are
needed.  The best dataset to make this crucial check is the redshift
survey of giant elliptical galaxies planned in the Sloan Digital Sky
Survey (SDSS, see astro-ph/9810130, astro-ph/9809085,
astro-ph/9809179, astro-ph/9805314).  Giant elliptical galaxies are
concentrated to clusters and superclusters, and their distribution on
large scales provides the best test for the shape of the power
spectrum of matter on scales around the maximum.  Presently we can say
that the possibility of a broken-scale-invariant primordial power
spectrum deserves serious attention.

\noindent We thank U. Seljak and M. Zaldarriaga for permission to use
their CMBFAST package, M. Gramann for permission to use the program of
the Press-Schechter algorithm, S. Bonometto, E. Gawiser, A. Szalay,
and D. Tucker for discussion, and H. Andernach for his help to improve
the style of the text.  We thank the referee, M. Vogeley, for
constructive criticism.  This work was supported by Estonian Science
Foundation grant 2625, and International Science Foundation grant
LLF100. JE and AS were supported by the Deutsche
Forschungsgemeinschaft in Potsdam; AS was partially supported by the
Russian Foundation for Basic Research under Grant 96-02-17591; RC was
supported by grants NAG5-2759, AST9318185.


\section*{Appendix: Cluster mass function}

\begin{figure}[ht]
\vspace*{6.5cm}  
\caption{Cluster mass functions. Open and filled circles with error
bars indicate the observed cluster mass function according to Bahcall
and Cen (1993); solid and dashed lines show mass functions calculated
for the linear empirical power spectra, $P_{HD}(k)$ and $P_{MD}(k)$,
for density parameters $\Omega_0=1.0, ~0.6, ~0.4$.  }
\includegraphics{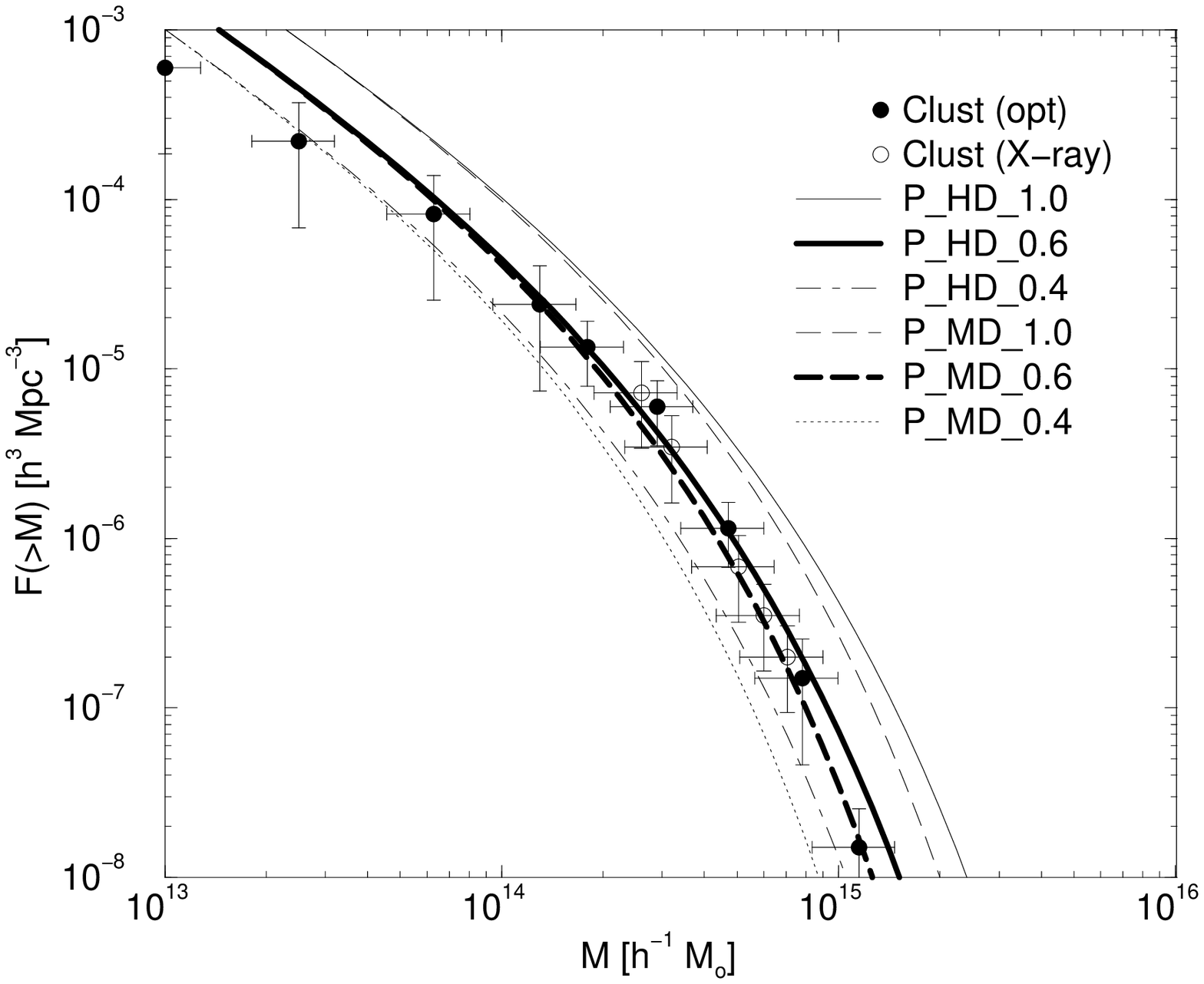}
\label{figure5}
\end{figure}

We use the empirical power spectra of mass to calculate the cluster
mass function. The latter can be compared with the observed cluster
mass function which allows us to check both sets of data for
consistency.

We calculated the cluster mass function using the algorithm, developed
by Suhhonenko \& Gramann (1998) on the basis of the Press-Schechter
(1974) theory. This function provides an important constraint for the
power spectrum (White, Efstathiou \& Frenk 1993, Cen 1998). The
cluster mass function was found for both empirical power spectra,
$P_{HD}(k)$, and $P_{MD}(k)$, reduced to the linear regime.  Mass
functions were derived for three values of the density parameter,
$\Omega_0=1.0, ~0.6, ~0.4$. Results are shown in Figure~5, together
with the observed cluster mass function according to Bahcall \& Cen
(1993). Our calculations show that both variants of the empirical
power spectra yield a cluster mass function which has a shape close to
the shape of the observed cluster mass function. The differences
between mass functions for spectra $P_{HD}(k)$ and $P_{MD}(k)$ are
small, in other words, the mass function is not sensitive to such
differences of power spectra. This is not surprising since clusters of
galaxies are formed by density perturbations on 1 -- 10 \Mpc\ scales
where both empirical power spectra are identical. On the other hand,
the amplitude of the mass function is very sensitive to the density
parameter of the Universe used for calculations. The amplitude is
evidently too high if one assumes a high density parameter of the
Universe. The best agreement is obtained for the case of
$\Omega_0=0.6$. A value of $\Omega_0 = 0.4$, as favored by cluster
abundance evolution constraints (Bahcall \& Cen 1993, Bahcall, Fan \&
Cen 1997), yields a lower amplitude of the calculated cluster mass
function.

The focus of this series of papers is on the empirical power spectrum
and on its consequences for the primordial power spectrum. We do not
consider the calculation of the cluster mass function for empirical
power spectra as a density determination.  The only conclusion we can
draw from this comparison is that our power spectra yield cluster mass
functions in good agreement with the observed cluster mass function. A
relatively high value of the density parameter favored by this test can be
explained if we assume that the biasing correction found from the
fraction of matter in high-density regions is too large (see Paper
II).



\end{document}